\documentclass{article}
\usepackage{amssymb}
\usepackage{authblk}
\usepackage{amsmath}
\usepackage{latexsym}
\usepackage{array}
\usepackage{graphicx}
\usepackage{bm}
\usepackage{exscale}
\usepackage{xcolor}
\usepackage{overpic}
\usepackage{booktabs}
\usepackage{tikz}
\usepackage{epic,eepic}
\usepackage{comment}
\usepackage[normalem]{ulem}
\usepackage{pxrubrica}
\usepackage{multirow}
\newlength\savedwidth

\usepackage{mathtools}
\newcommand{\keshi}[1]{\begin{color}{blue!40}\sout{#1}\end{color}}
\newcommand{\tsuika}[1]{\begin{color}{orange}{#1}\end{color}}

\usepackage{theorem}
\theoremstyle{break}

\def\Z{{\mathbb Z}}

\def\R{{\mathbb R}}

\makeatletter 
  \renewcommand{\@biblabel}[1]{#1.}
  \makeatother
\begin{document}
\title{Pattern formation of elliptic particles by two-body interactions: a model for dynamics of endothelial cells in angiogenesis} 
%
\author[1]{Tatsuya Hayashi}
\author[2]{Fumitaka Yura}
\author[3]{Jun Mada}
\author[4]{Hiroki Kurihara}
\author[5]{Tetsuji Tokihiro}

\affil[1]{\small Graduate School of Information Science and Technology, Hokkaido University, Kita 14, Nishi 9, Kita-ku, Sapporo, Hokkaido, 060-0814, Japan.}
\affil[2]{\small School of Systems Information Science, Future University Hakodate, 116-2 Kamedanakano-cho, Hakodate, Hokkaido, 041-8655, Japan.}
\affil[3]{\small College of Industrial Technology, Nihon University, 1-2-1, Izumi-cho, Narashino, Chiba, 275-8575, Japan.}
\affil[4]{\small Graduate School of Medicine, the University of Tokyo, 7-3-1, Hongo, Bunkyo-ku, Tokyo, 113-0033, Japan.}
\affil[5]{\small Graduate School of Mathematical Science, the University of Tokyo, 3-8-1, Komaba, Meguro-ku, Tokyo, 153-8914, Japan.}
\date{}
\maketitle

\begin{abstract} %
A two-dimensional mathematical model for dynamics of endothelial cells in angiogenesis is investigated.
Angiogenesis is a morphogenic process in which new blood vessels emerge from an existing vascular network. 
Recently a one-dimensional discrete dynamical model has been proposed  to reproduce elongation, bifurcation, and cell motility such as cell-mixing during angiogenesis on the assumption of a simple two-body interaction between endothelial cells. 
The present model is its two-dimensional extension, where endothelial cells are represented as the ellipses with the two-body interactions
: repulsive interaction due to excluded volume effect, attractive interaction through pseudopodia and rotation by contact. 
We show that the oblateness of ellipses and the magnitude of contact rotation significantly affect the shape of created vascular patterns and elongation of branches. 
\end{abstract}
%
%
%
\section{Introduction}

Collective cell migration is the coordinated multicellular movement 
in response to interactions with the 
extracellular matrix (ECM) and other cells~\cite{haeger2015,hagiwara2021}. Collective motion is an important factor for morphogenesis and is widely observed in both physiological and pathological 
processes, such as blood vessel sprouting, cancer metastasis, and tissue repair~\cite{omelchenko2003,friedl2003,scarpa2016,zhang2019}. 
In particular, the formation of blood vessels is one of the fundamental phenomena in morphogenic process~\cite{potente2017,vaahtomeri2017}. 
Vascular morphogenesis occurs primarily through sprouting angiogenesis, a process in which new blood vessels emerge from an existing vascular network~\cite{eilken2010,ribatti2012,fonseca2020}. Although a collective migration of endothelial cells (ECs) contributes to sprout elongation and branch formation, single cell imaging technique has revealed that ECs exhibit quite complex movement, such as moving forwards and backwards, overtaking~\cite{jakobsson2010,arima2011} and so on. The question of how ECs form branch structures with their heterogeneous motions is one of the most interesting subjects, both experimentally and theoretically. 

Theoretical models for angiogenic sprouting have been extensively studied perspective in various contexts such as pathology and physiology~\cite{bauer2007,daub2013,mada2016,perfahl2017,sasaki2017}. 
In our previous work, we proposed a discrete dynamic model for the dynamics of ECs during angiogenesis on the basis of \textit{in vitro} experiments~\cite{arima2011,sugihara2015}, and verified that the deterministic two-body interaction between ECs can bring about cell-mixing, elongation, and branching~\cite{matsuya2016}. 
In addition, we 
estimated parameters and confirmed validity of the model by comparing it with the movement of ECs on a two-dimensional plane obtained by \textit{in vitro} experiments using mouse aortic tissue~\cite{takubo2019}. The 
results suggested the existence of attractive force in an area about the length of a pseudopod and 
repulsive force 
which works over shorter distances. The attraction 
is a self-driven force originating from cell communication caused by the adhesion between proteins on the surface of the cell membrane, while the short-range repulsion 
is considered 
to be the exclusion volume effect. In general, the excluded volume effect depends on the shape, such as chains 
in protein folding, and rod-shape molecules 
in liquid crystals. Therefore, to understand the dynamics of ECs under this attractive and repulsive 
forces, a mathematical model considering the shape of ECs is required.

The relationship between the shape of ECs and the pattern formation has been well studied experimentally and theoretically. Experiments showed that vascular endothelial growth factor (VEGF) induces 
elongated 
shapes in ECs~\cite{drake2000}. 
A group of untreated ECs, formed a network formation, while ECs blocked VEGF signaling showed a round shape and formed aggregated clusters (Fig. 5 in~\cite{drake2000}). Mathematical models based on the cellular Potts model showed that cell elongation is important for network formation during vasculogenesis~\cite{merks2006,palm2013}. A cell-based model in which cells 
are represented by ellipses on a plane was proposed
. It showed that elongated ellipses with large aspect ratio formed network structures, and 
 the orderliness of the 
patterns was evaluated~\cite{palachanis2015}. 
The 
equation of motion of each cell was modeled 
by a Langevin equation, and the attractive or repulsive interaction 
was determined by the overlap area between two ellipses of two cells
. To minimize overcrowding and maximize attraction, the rotation of an ellipse is determined by using Monte Carlo 
methods with the acceptance probability depending on the overlap area and a noise parameter.

In this paper, we present a discrete mathematical model for the dynamics of ECs during angiogenesis considering the shape of an EC on the basis of the results in~\cite{takubo2019}. 
Using the proposed model, we investigate the 
relationship between the shape of an ellipse and pattern formation. We also examine how rotation of an ellipse affects collective motion 
and the elongation of branches.
Although we represent ECs by ellipses, an essential difference of the present model and that in Ref.~\cite{palachanis2015} is that the interactions 
in our model are only two-body interactions and are completely deterministic even for rotation of ECs.
%

%
%
\section{Elliptic particle model with two-body interaction}
Endothelial cells migrate by stretching their pseudopodia. It is reasonable to assume that the interaction of pseudopodia between cells generates self-driving forces in those cells. To incorporate the anisotropy of cell shape caused by the elongation of pseudopodia, we 
assumed that each cell is described as an ellipse on a plane. 
This section explains how to determine whether two ECs have collided or not. 
We give a mathematical model for the dynamics of elliptic particles. 
Then, we describe the settings in numerical simulation and parameters used in simulations.

\subsection{Collision of two elliptic particles}
Let $a_i$ and $b_i$ be the major and minor axes of the $i$-th cell (hereafter called ``cell-$i$''), respectively.
The two-dimensional coordinates of the boundary (ellipse) of the cell-$i$ is expressed as  
\begin{equation}
\begin{pmatrix}
x_i(\theta) \\
y_i(\theta) \\
\end{pmatrix}
= {\bm r}_i + 
\begin{pmatrix}
\cos \psi_i & -\sin \psi_i \\
\sin \psi_i & \cos \psi_i \\
\end{pmatrix}
\begin{pmatrix}
a_i & 0 \\
0  & b_i \\
\end{pmatrix}
\begin{pmatrix}
\cos \theta \\
\sin \theta \\
\end{pmatrix}\quad (0 \le \theta <2\pi),
\end{equation}
where ${\bm r}_i \in \R^2$ and $\psi_i \in [0, \pi)$ denote the position of center of gravity and inclination of the major axis of cell-$i$, respectively. 
For $D \in \Z_{>0}$ and $\theta_k := 2\pi k/D$, we take $D$ sampling points on the ellipse of cell-$i$ as ${}^t(x_i(\theta_k),y_i(\theta_k))$ $(k=0, 1, \dots, D-1)$. We regard that cell-$i$ and cell-$j$ 
collide with each other when some sampling points on the ellipse of cell-$i$ locate inside the ellipse of cell-$j$, and vice versa. Since an ellipse is given by a quadratic equation, we can easily check whether the sampling points are inside an ellipse or not.

\subsection{The dynamics of elliptic particles}
\begin{figure}[h]
 \centering
 \begin{overpic}[width=15.0cm]{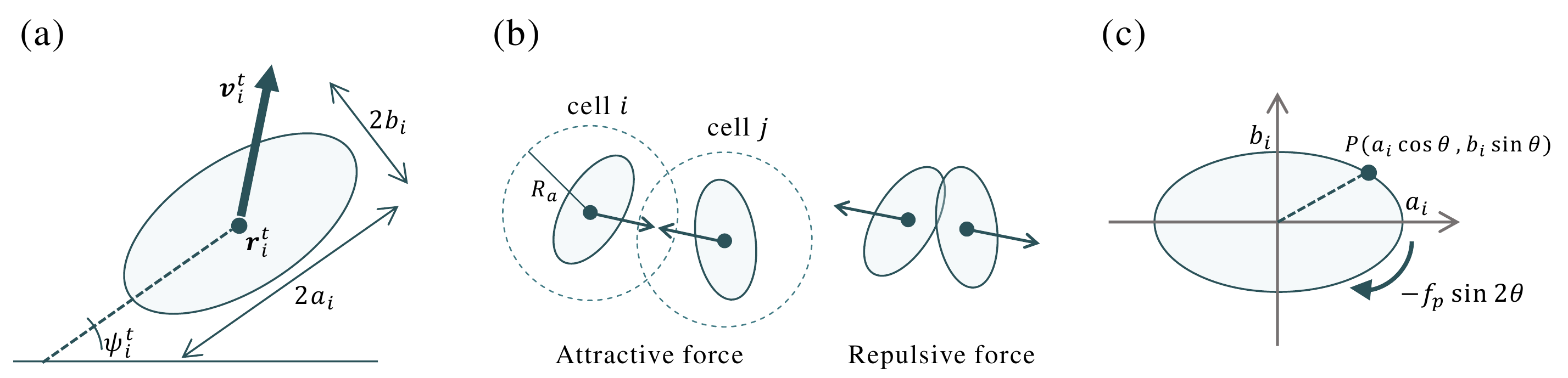}
 \end{overpic}
 \caption{Illustration of our proposed mathematical model. (a) ECs are described as 2D elliptic particles. (b) Two-body interaction between particles: attractive force (left) and repulsive force (right). The dashed circle of radius $R_a$ shows the interaction domain of cell-$i$. (c) Rotational force. The horizontal axis refers to the direction of the major axis, and the vertical axis refers to the direction of the minor axis. According to a point of contact $P$ on the ellipse, the force $-f_p \sin 2\theta$ acts on a particle.}
 \label{fig:model}
\end{figure}

The state of cell-$i$ at time step $t \in \Z_{t \geq 0}$ is characterized by its position ${\bm r}_i^t \in \R^2$, velocity ${\bm v}_i^t \in \R^2$ and inclination of major axis $\psi_i^t \in [0, \pi)$ (Fig.~\ref{fig:model} (a)). We consider the following discrete dynamical systems\keshi{.}\tsuika{:}

\begin{align}
{\bm r}_i^{t+1} &= {\bm r}_i^t + {\bm v}_i^t, \label{eq:position} \\
{\bm v}_i^{t+1} &= {\bm v}_i^t - \gamma {\bm v}_i^t + \sum_{j \neq i} {\bm F}_{i,j}^t, \label{eq:velocity} \\
\psi_i^{t+1} &= \psi_i^t - \sum_k{}^\prime f_p \sin 2\theta_k, \label{eq:angle}
\end{align}
where the parameter $\gamma >0$ denotes the coefficient of conflict, ${\bm F}_{i,j}^t$ denotes the two-body interaction, $f_p$ is a positive constant, and $\sum_k^{\prime}$ denotes the summation over all sampling points $\theta_k$ of cell-$i$ contained inside other ellipses. 

For the force ${\bm F}_{i,j}^t$, we 
adopt the following form: 
\begin{itemize}
\item[(i)] If sampling points on the ellipse of the cell-$i$ are in that of the cell-$j$, then ${\bm F}_{i,j}^t = f_r {\bm e}_{i,j}^t$
\item[(ii)] Else, if $\| {\bm r}_i^t-{\bm r}_j^t  \| \leq R_a$, then ${\bm F}_{i,j}^t = - f_a {\bm e}_{i,j}^t$
\item[(iii)] Otherwise ${\bm F}_{i,j}^t = \boldsymbol{0}$.
\end{itemize}
Here, $R_a$ is the threshold 
length for the attraction, 
and $f_a$, $f_r$ are the positive constants for the strength of interaction (Fig.~\ref{fig:model}(b)). 
The vector ${\bm e}_{i,j}^t$ is defined as
\begin{equation}\label{eq:unitvector}
{\bm e}_{i,j}^t := \left(  {\bm r}_i^t-{\bm r}_j^t \right)/\|  {\bm r}_i^t-{\bm r}_j^t  \|.
\end{equation}

Equations~\eqref{eq:position} and \eqref{eq:velocity} show the discrete analogue of the Newtonian equation of motion. Equation~\eqref{eq:angle} means that when two elliptic particles come into contact, their major axes rotate in the direction to avoid the collision (Fig.~\ref{fig:model} (c)). The parameter $f_p$ in 
Eq.~\eqref{eq:angle} controls the orientation of ellipses.  The particles do not rotate at all if $f_p = 0$, and they become to rotate easier as the value $f_p$ increases.

\subsection{The settings in numerical simulation}
Let us give the values of parameters used in numerical simulations. The parameters for an ellipse 
are set to be the same for all particles ($\forall i, a_i=a, b_i=b$), and the area 
is scaled to one, i.e.
, $ab=1$. We use 
an oblateness $f:=1-b/a$ as a parameter for the shape of an ellipse. For a given $f$, $a$ and $b$ are determined from $ab=1$ and $1-b/a=f$. Reference values of other parameters were $D=16$, $R_a = 1.5 a$, $\gamma = 0.1$, $f_a = 0.002$, $f_r = 0.05$, and $f_p = 0.005$. 
Comparing with experimental situations~\cite{takubo2019}, $ab \sim 2500\, \mathrm{\mu m}^2$, which implies that unit length is about $50\, \mathrm{\mu m}$, and one time step of Eqs.~(\ref{eq:position})-(\ref{eq:angle}) is about $10$ minutes.

In numerical simulations, we 
treat a square domain $[-30, 30] \times [-30, 30]$ and 
examine two situations. One is the case where $500$ particles are initially distributed at random in the domain and no particle is supplied in time. The other is that only one particle exists initially at the origin and a particle is supplied to the origin every ten time steps. We shall refer to the former as the case of random distribution and the latter as that of constant supply. The initial velocity and orientation angle of a particle 
are randomly selected  in the both cases according to the uniform distribution over $[-0.1, 0.1] \times [-0.1, 0.1]$ and that over $[0, \pi)$, respectively. The patterns obtained after $5000$ time steps are characterized by using their fractal dimensions.

As for the boundary conditions, we 
consider three kinds of boundary conditions: a free boundary condition, a periodic boundary condition, and a wall (Dirichlet) boundary condition. 
We mainly 
use a free boundary condition. To investigate the influence of boundary conditions, we 
examine periodic and wall boundary conditions in Sec.~\ref{sec:boundary}.

%
%
\section{Results and Discussions}

\subsection{A small number of particles with larger oblateness align linearly}

\begin{figure}[h]
 \centering
 \begin{overpic}[width=15.0cm]{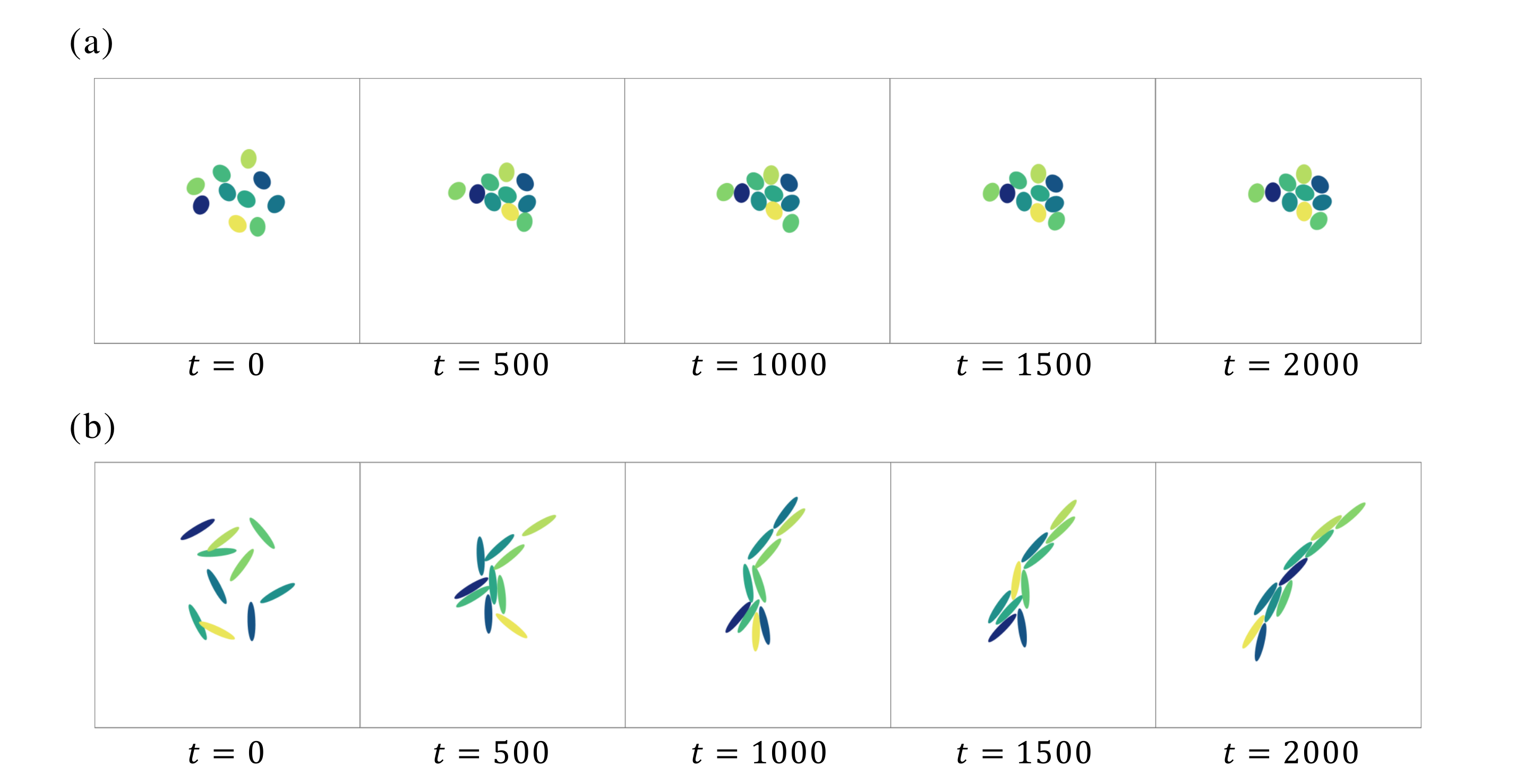}
 \end{overpic}
 \caption{Snapshots of numerical simulation for ten particles with small and large oblateness: (a) $f=0.2$, (b) $f=0.8$. All parameters except for the oblateness are the same in (a) and (b): $\gamma = 0.1$, $f_a = 0.002$, $f_r = 0.05$, and $f_p = 0.01$.}
 \label{fig:10cells}
\end{figure}

We show some examples of numerical simulation for small number of particles. Simulations 
are initialized with ten particles randomly distributed on a square domain, and their initial velocities 
are all set to be zero. 
Figures~\ref{fig:10cells} (a) and (b) show time evolution for $f=0.2$ and $0.8$, respectively. 
Elliptic particles with small oblateness ($f=0.2$) aggregate and 
barely move (Fig.~\ref{fig:10cells} (a)). 
On the other hand, elliptic particles with large oblateness ($f=0.8$) line up in a straight line 
move actively so that the direction of alignment changes in time (Fig.~\ref{fig:10cells} (b)). The reason 
for this behavior is that when elongated elliptic particles contact, they tend to rotate so that their major axes are aligned. We see that the shape of a particle significantly affects the particle alignment even for small number of particles.

%

To see how often particles change their position, we consider the following quantity:
\begin{equation}
w_R^{T_0} (t) := \frac{1}{T_0 N(N-1) \pi} \sum_{i < j} \left| \varphi_{i,j}^{t+T_0} - \varphi_{i,j}^{t} \right|,
\end{equation}
where $T_0$ is a positive integer, $N$ is the total number of particles, and $\varphi_{i,j}^t$ is the angle of ${\bm e}_{i,j}^t$ in Eq.~(\ref{eq:unitvector}), that is, ${\bm e}_{i,j}^t = {}^t\left(\cos \varphi_{i,j}^t, \sin \varphi_{i,j}^t \right)$. Then, we define the average winding number $w_R^{T_0}$ as follows:
\begin{equation}
w_R^{T_0} := \langle w_R^{T_0} (t) \rangle = \lim_{m \to \infty} \frac{1}{m} \sum_{k=1}^{m} w_R^{T_0}(k T_0).
\end{equation}

\begin{figure}[h]
 \centering
 \begin{overpic}[width=13.0cm]{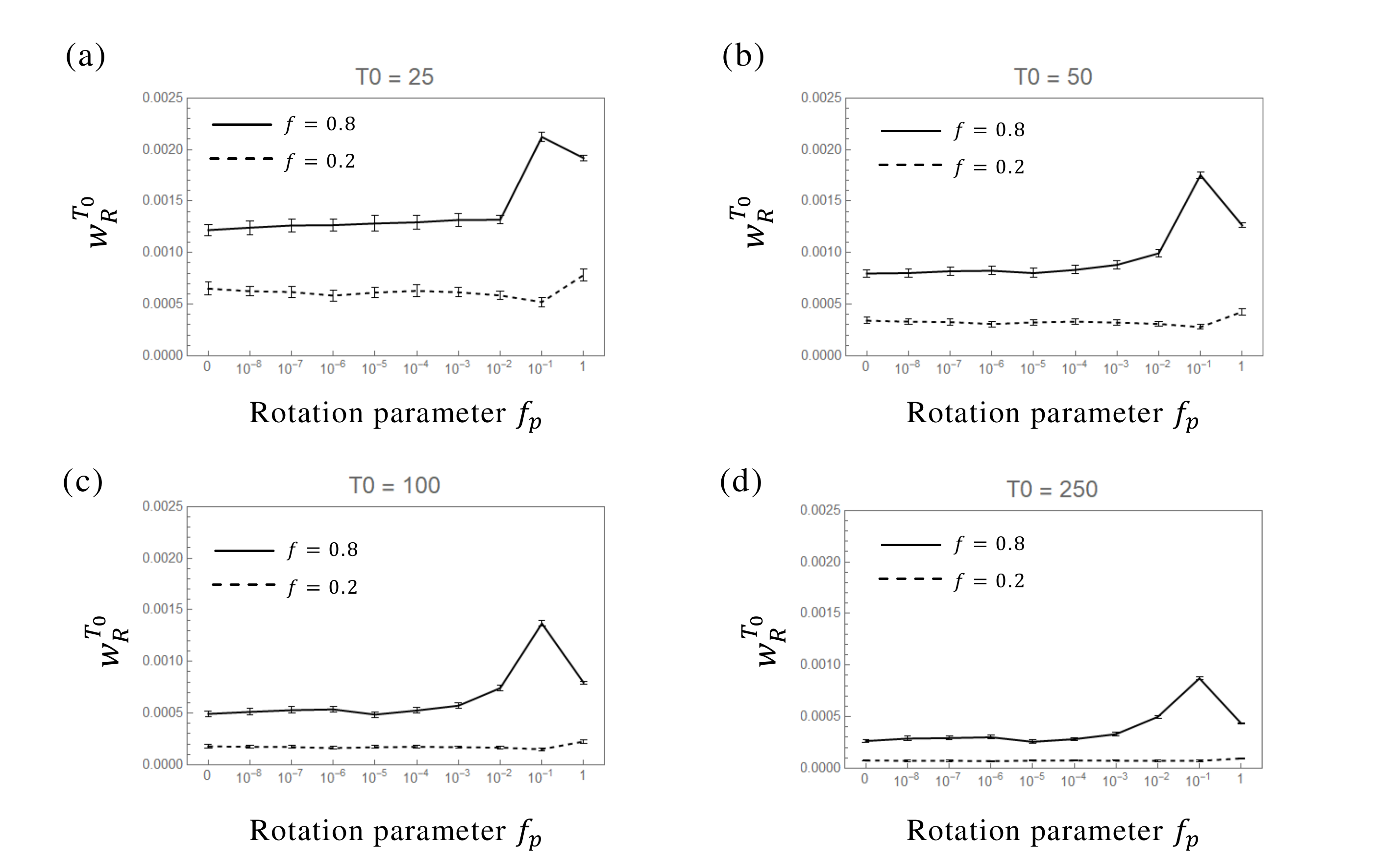}
 \end{overpic}
 \caption{Relationship between the average winding number $w_R^{T_0}$ and the parameter of rotation $f_p$ for (a) $T_0 = 25$, (b) $T_0 = 50$, (c) $T_0 = 100$ and (d) $T_0 = 250$. The horizontal axis is $f_p$ and the vertical one is the number of rotations. These plots show the average over 100 simulations (error bars are 95\% confidence interval).}
 \label{fig:winding}
\end{figure}

Figures~\ref{fig:winding} show $w_R^{T_0}$ for $T_0 = 25, 50, 100$ and $250$. We have a sharp peak at $f_p=0.1$ in all the cases. Thus, we see that rotation of particles is more pronounced when the value $f_p$ is of appropriate magnitude.

\subsection{Pattern formation for large system size}
\begin{figure}[h]
 \centering
 \begin{overpic}[width=15.0cm]{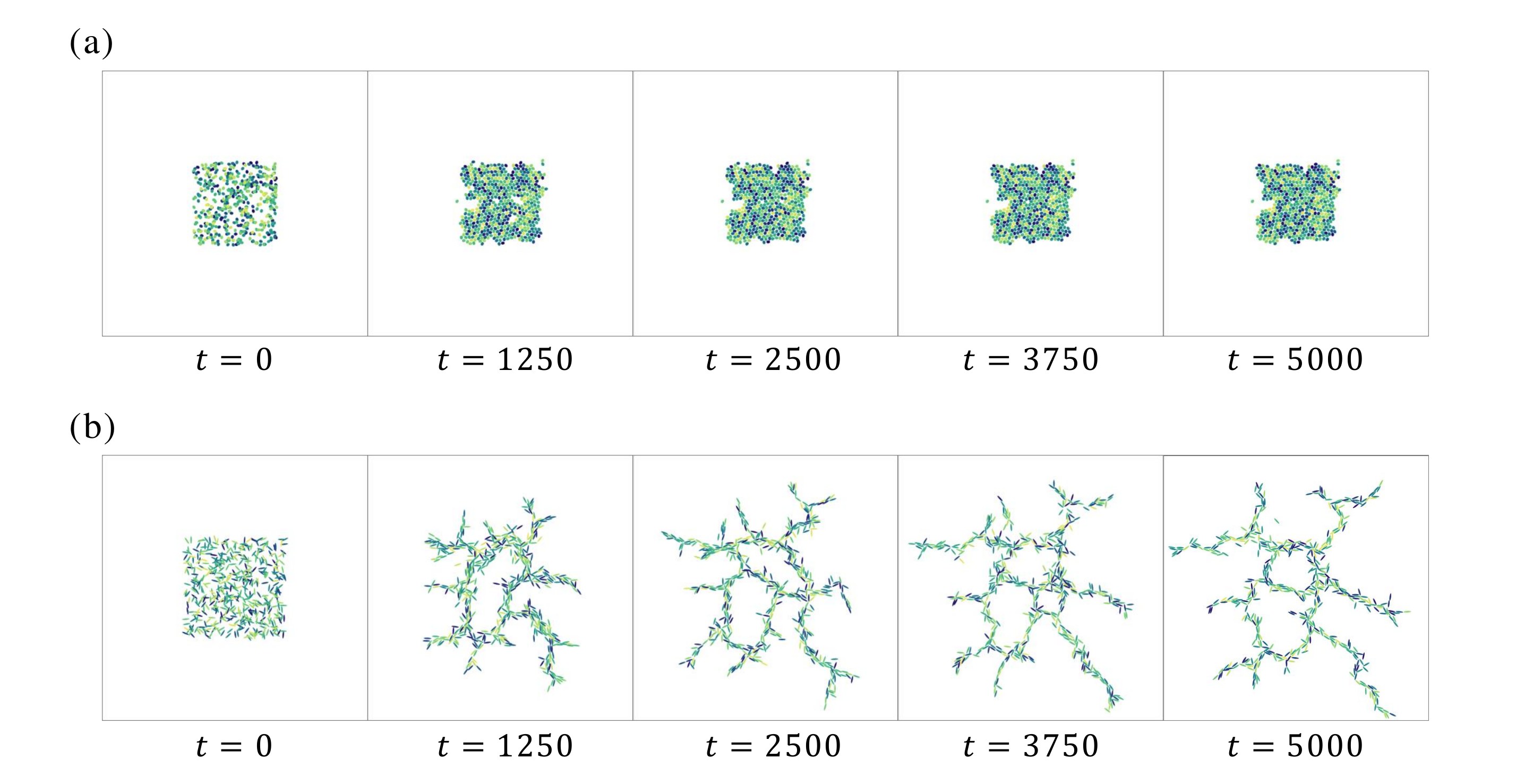}
 \end{overpic}
 \caption{Snapshots of numerical simulation for $500$ particles with small and large oblateness: (a) $f=0.2$, (b) $f=0.8$. All parameters except for the oblateness are the same in (a) and (b): $\gamma = 0.1$, $f_a = 0.002$, $f_r = 0.05$, and $f_p = 0.005$. Elliptic particles are randomly distributed at $t=0$.}
 \label{fig:500random}
\end{figure}

\begin{figure}[h]
 \centering
 \begin{overpic}[width=15.0cm]{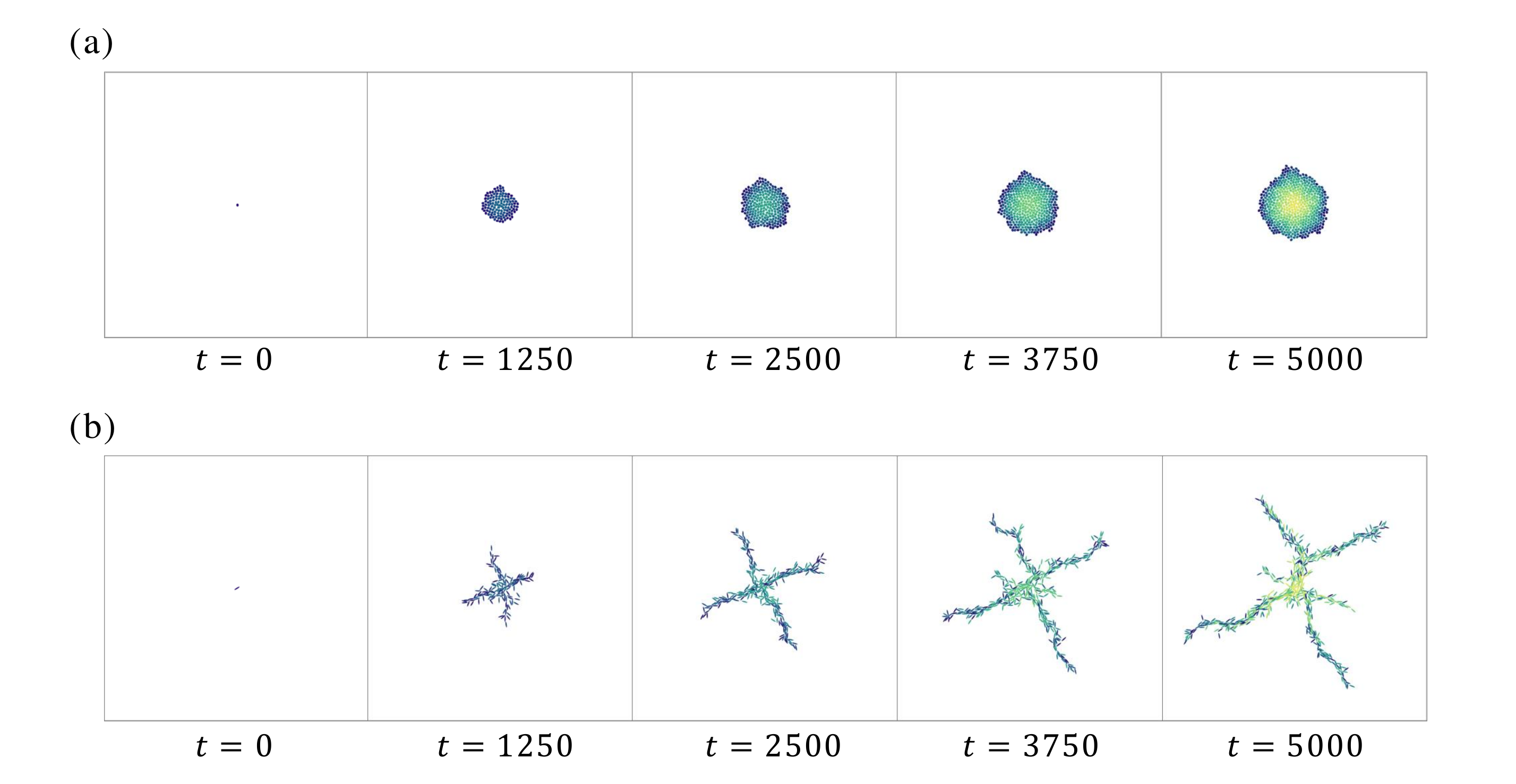}
 \end{overpic}
 \caption{Snapshots of numerical simulation for $500$ particles with small and large oblateness: (a) $f=0.2$, (b) $f=0.8$. All parameters except for the oblateness are the same in (a) and (b): $\gamma = 0.1$, $f_a = 0.002$, $f_r = 0.05$, and $f_p = 0.005$. At $t=0$, single particle is placed at the origin, and a particle is supplied to the origin every ten steps.}
 \label{fig:500supply}
\end{figure}

We show examples of the time evolution of $500$ elliptic particles with oblateness $f=0.2$ and $0.8$. Round particles ($f=0.2$) always aggregate in the case of random distribution (Fig.~\ref{fig:500random} (a)) and that of constant supply (Fig.~\ref{fig:500supply} (a)). In contrast, elongated particles form network structures observed in vasculogenesis (Fig.~\ref{fig:500random} (b)), and branch-like structures observed in angiogenesis (Fig.~\ref{fig:500supply} (b)). In both cases, we obtain network or branch-like structures for elongated elliptic particles with $f=0.8$.

\begin{figure}[h]
 \centering
 \begin{overpic}[width=15.0cm]{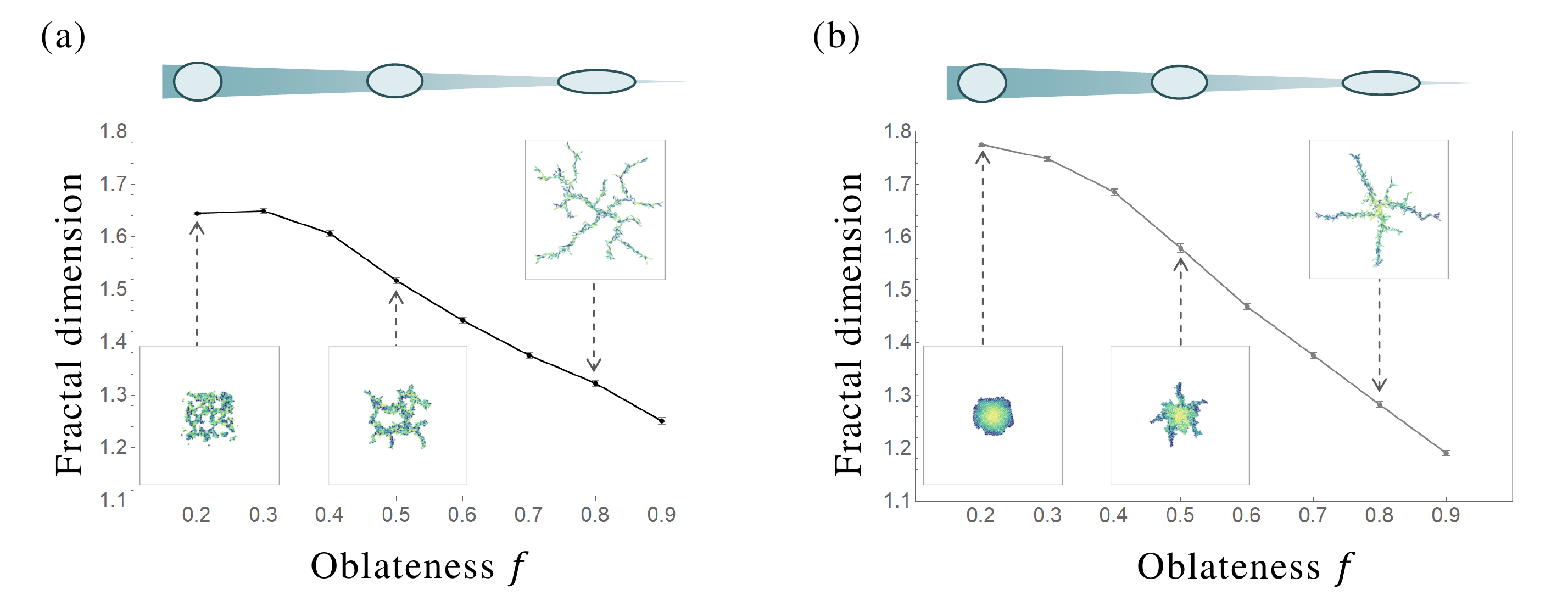}
 \end{overpic}
 \caption{The dependency of the shape of the ellipse and collective motion pattern for two types of simulations: (a) Elliptic particles are randomly distributed within a square at $t=0$. (b) Elliptic particle is supplied every ten steps to the origin. The horizontal axis shows the oblateness $f$ and the vertical one the fractal dimension. The figures in the graphs show snapshots at $t=5000$ for $f=0.2, 0.5, 0.8$. Common parameters are $\gamma = 0.1$, $f_a = 0.002$, $f_r = 0.05$, and $f_p = 0.005$. The plots are averaged over $100$ simulations and their $95$\% confidence intervals.}
 \label{fig:oblateness}
\end{figure}

To characterize the patterns shown in Figs.~\ref{fig:500random} and \ref{fig:500supply}, we 
compute the fractal dimensions of the network structures. The oblateness $f$ is set from 
$0.2$ to $0.9$ in increments $0.1$. Figure~\ref{fig:oblateness} shows the relationship between the oblateness and the fractal dimension. In the case of random distribution, the fractal dimension decreases as the oblateness increases (Fig.~\ref{fig:oblateness} (a)). The fractal dimension 
for the case of small oblateness takes a 
larger value because elliptic particles with small oblateness aggregate and spread out on a plane. Conversely, the fractal dimension of a pattern formed by particles with the large oblateness takes a 
smaller value. In the case of constant supply, the fractal dimension also decreases as the oblateness increases (Fig.~\ref{fig:oblateness} (b)). 
The pattern for small oblateness is almost two-dimensional area, while that for large oblateness exhibits one-dimensional branching structures.

%

\begin{figure}[h]
 \centering
 \begin{overpic}[width=15.0cm]{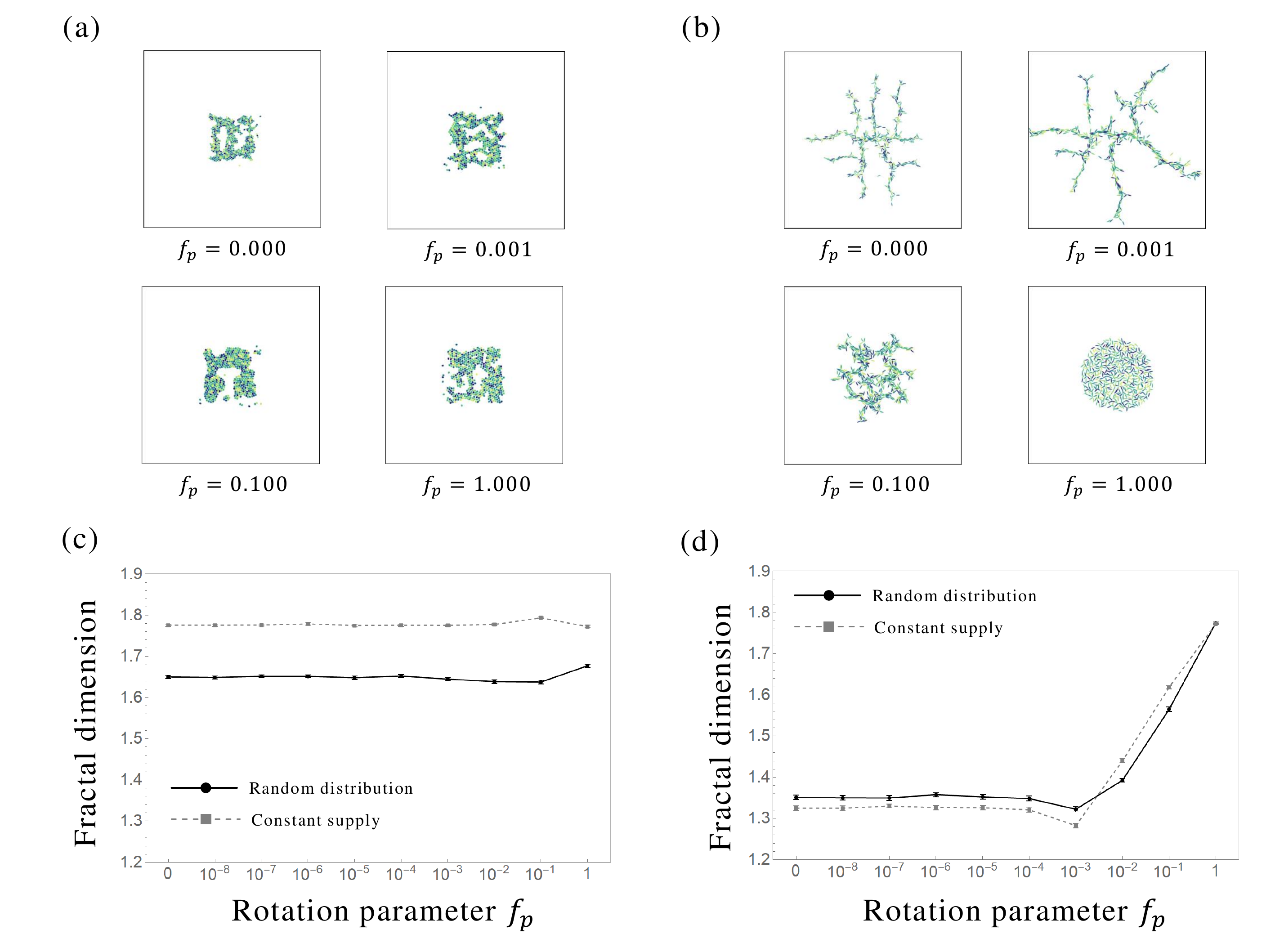}
 \end{overpic}
 \caption{The relationship between the rotation of an elliptic particle and collective motion pattern. The panels (a) and (b) show examples of simulation for oblateness $f=0.2, 0.8$ and the parameter of rotation $f_p=0.000, 0.001, 0.100$ and $1.000$ when elliptic particles are randomly placed as the initial configuration: (a) $f=0.2$, $f_p=0.000, 0.001, 0.100$ and $1.000$, (b) $f=0.8$, $f_p=0.000, 0.001, 0.100$ and $1.000$. (c) The dependency of aggregation and rotation for $f=0.2$. (d) The dependency of aggregation and rotation for $f=0.8$. The horizontal axis shows the parameter of rotation $f_p$ and the vertical one the fractal dimension. Common parameters are $\gamma = 0.1$, $f_a = 0.002$, and $f_r = 0.05$. The plots of (c) and (d) are averaged over $100$ simulations and their $95$\% confidence intervals.}
 \label{fig:rotation}
\end{figure}

Figures~\ref{fig:rotation} show the effect of the rotation on patterns for the oblateness $f=0.2$ and $0.8$. Since round ellipses ($f=0.2$) are less susceptible to rotation, the rotation parameter $f_p$ has no effect on patterns and particles are only uniformly distributed (Fig.~\ref{fig:rotation} (a)). The fractal dimension is comparable regardless of $f_p$ (Fig.~\ref{fig:rotation} (c)). Particles with large oblateness ($f=0.8$) form 
different patterns strongly depending on the rotation parameter $f_p$ from branch-like structures to uniformly aggregated structures (Fig.~\ref{fig:rotation} (b)). The fractal dimension remains almost unchanged for small $f_p$, and increases from $f_p=10^{-3}$ as shown in Fig.~\ref{fig:rotation} (d). The fractal dimension takes the minimum value around $f_p=10^{-3}$, which suggests that moderate rotation helps to align ellipses linearly and form branch-like structures. When the rotation parameter becomes larger, elongated particles become more sensitive to their contact and rotate quickly in collision. Hence, the strong rotational effect prevents particles from aligning linearly, and no network structure is created. 

\subsection{Moderate rotation can elongate branches}
\begin{figure}[h]
 \centering
 \begin{overpic}[width=11.0cm]{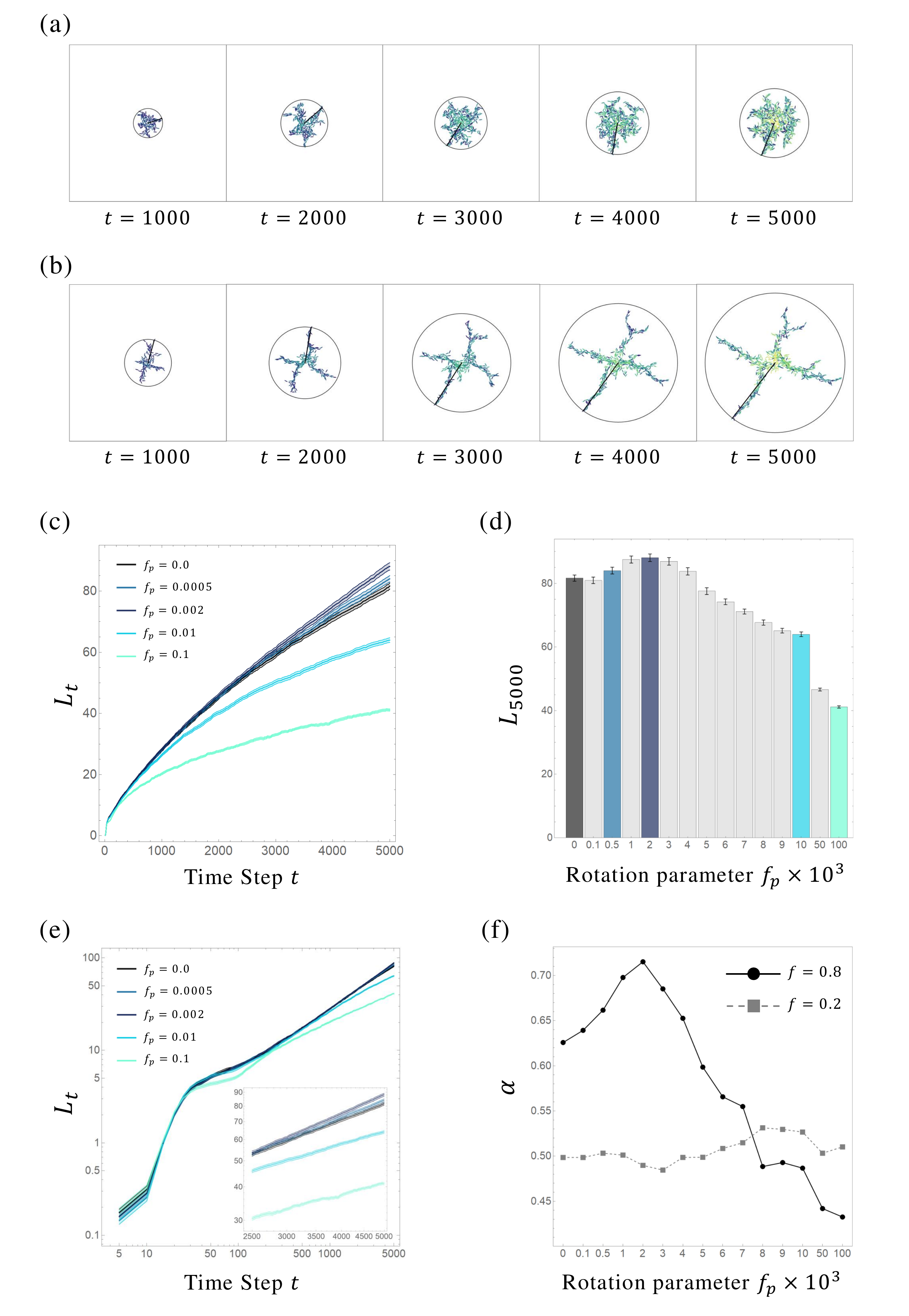}
 \end{overpic}
 \caption{Effect of rotation parameter $f_p$ on branch elongation. The top figures (a) and (b) show snapshots of the patterns for $f_p=0.1$ and $f_p=0.002$, respectively. The black circles represent circles with centre at the origin and radius $L_t$. (c) Temporal change of $L_t$ for $f_p=0$, $0.0005$, $0.002$, $0.01$, and $0.1$. These curves and bands in (c) show the average over $100$ simulations and $95$\% confidence intervals, respectively. (d) $L_t$ at $t=5000$ for rotation parameters, and these plots are averaged over $100$ simulations and their $95$\% confidence intervals. (e) Log-log plot of the curves in (c). (f) The slope of line from $t=2500$ to $5000$ in (e). Common parameters are $f=0.8$, $\gamma = 0.1$, $f_a = 0.002$, and $f_r = 0.05$.}
 \label{fig:radius}
\end{figure}

\begin{figure}[h]
 \centering
 \begin{overpic}[width=11.0cm]{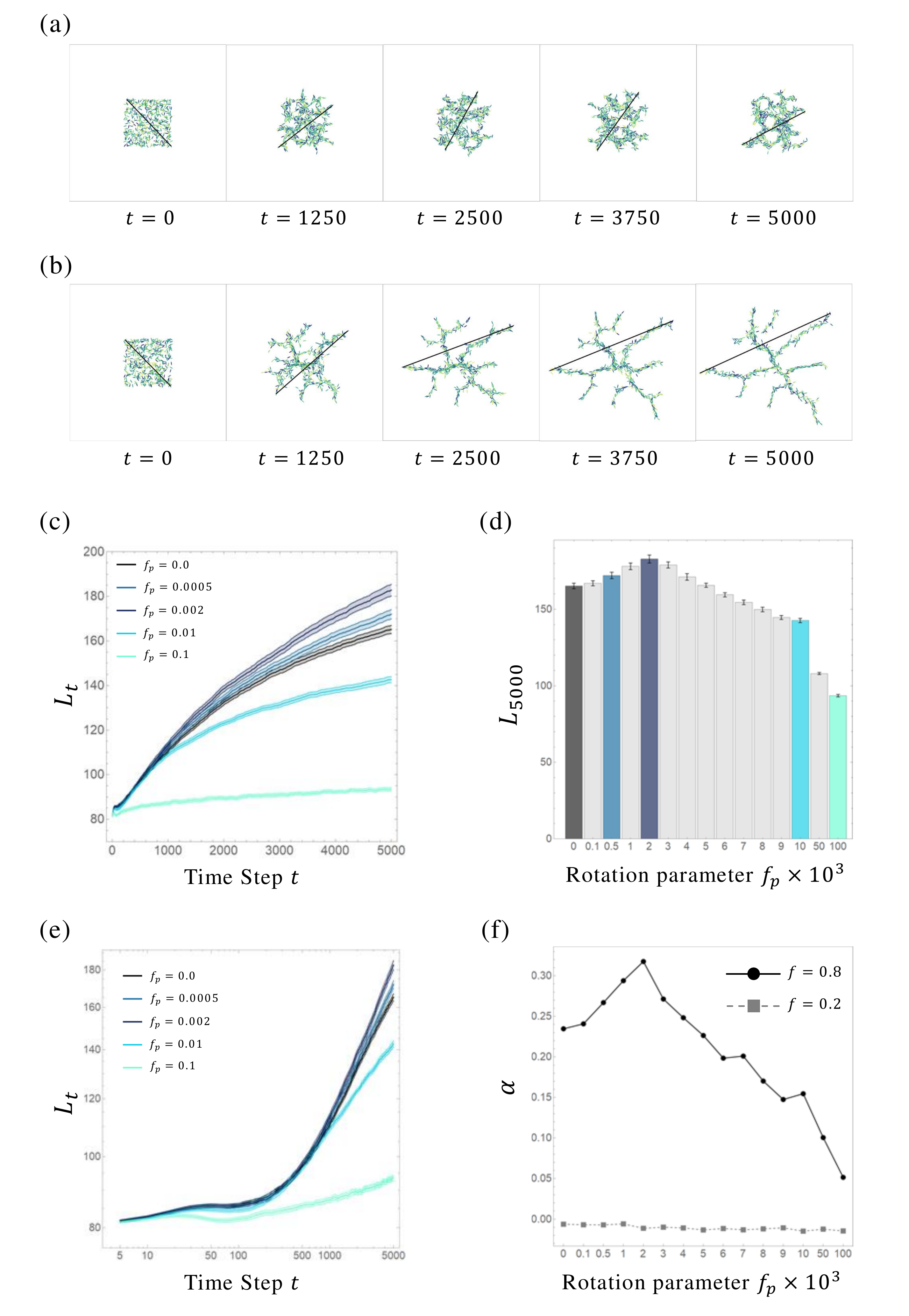}
 \end{overpic}
 \caption{Effect of rotation parameter $f_p$ on the diffusion of particles. The top figures (a) and (b) show snapshots of the patterns for $f_p=0.1$ and $f_p=0.002$, respectively. The black line shows the maximum cell-to-cell distance $L_t$ at time $t$. (c) Temporal change of $L_t$ for $f_p=0$, $0.0005$, $0.002$, $0.01$ and $0.1$. There curves and bands in (c) show the average over 100 simulations and their 95\% confidence intervals, respectively. (d) $L_t$ at $t=5000$ for rotation parameters, and these bars are averaged over 100 simulations (error vars are 95\% confidence intervals). (e) Log-log plot of the curves in (c). (f) The slope of line from $t=2500$ to $5000$ in (e). Common parameters are $f=0.8$, $\gamma=0.1$, $f_a=0.002$ and $f_r=0.05$.}
\label{fig:radius_random}
\end{figure}

We investigate how the rotation of a particle affects the elongation of a branch. 
Figures~\ref{fig:radius} 
show the influence of rotation parameter $f_p$ on the spread of a pattern for particles with the oblateness $f=0.8$. Here we consider the case of constant supply. Let $L_t$ be the maximum reaching distance of a particle 
from the supply point (the origin) at time $t$, i.e. $L_t = \max_i \| {\bm r}_i^t \|$. Figures~\ref{fig:radius} (a) and \ref{fig:radius} (b) show time-sequences of patterns for the large and small values of the parameter $f_p$, respectively. For large value of rotation parameter, patterns do not spread easily because quick rotation inhibits the formation of branches. Particles with sufficiently small rotation effect 
construct branches, and form a spread pattern by the elongation of branches. Figure~\ref{fig:radius} (c) shows the time variation of $L_t$ for $f_p=0$, $0.0005$, $0.002$, $0.01$, and $0.1$. The distance $L_t$  at $t=5000$ has maximum value around $f_p=0.002$ as shown in Fig.~\ref{fig:radius} (d). In Fig.~\ref{fig:radius} (c), for sufficiently large $t$, $L_t$ approximately increases according to a power law, i.e. $L_t \propto t^{\alpha} (\alpha \in \R)$. Figure~\ref{fig:radius} (e) shows the growth curve in Fig.~\ref{fig:radius} (c) on log-log scale.  For several values of rotation parameters $f_p$, we compute the exponent $\alpha$ by the time evolution of $L_t$ from $t=2500$ to $5000$ using the least-squares method (Fig.~\ref{fig:radius} (f)). The exponent $\alpha$ is maximal at $f_p=0.002$ for elongated particle ($f=0.8$), which suggests that moderate rotation is required for elongating branches. For round particles ($f=0.2$), $\alpha$ is around $0.5 \sim 0.55$ regardless of strengths of rotation parameter, as shown in Fig~\ref{fig:radius} (f). 
%
Figures~\ref{fig:radius_random} show the same graphs in random distribution. Here $L_t$ denotes the maximum distance between particles. The features of patterns are similar to those in constant supply, which means that these features with respect to rotation parameter do not depend on how to supply the particles.

%
%

\subsection{Effects of cell-mixing}
\begin{figure}[h]
 \centering
 \begin{overpic}[width=16.0cm]{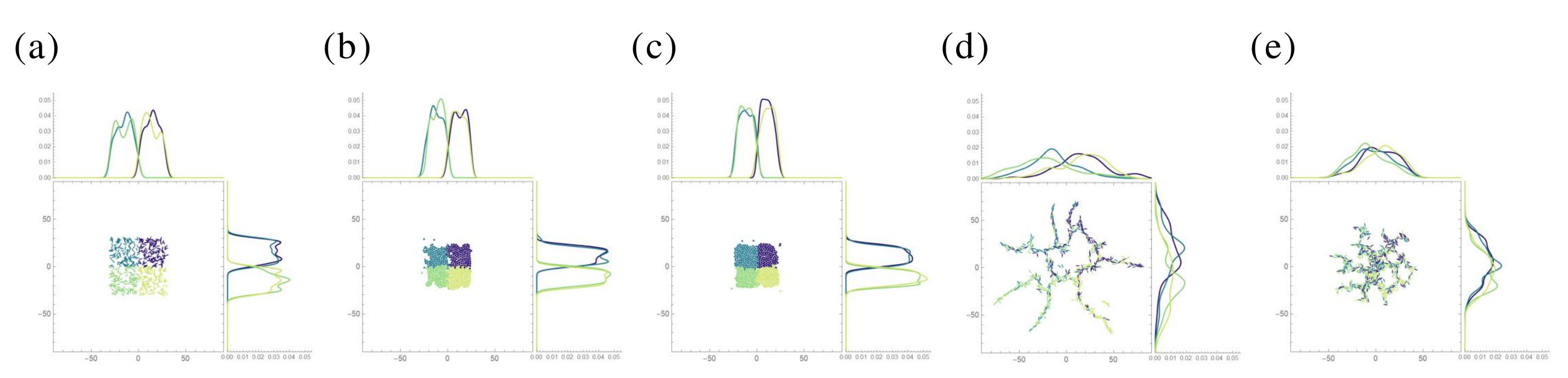}
 \end{overpic}
 \caption{Effect of oblateness $f$ and the parameter of rotation $f_p$ on pattern. (a) Particles are distributed at random within a square domain at $t=0$. Particles are color-coded into four colors according to their initial position. For each color, the top and right distributions show the probability density of $x$- and $y$-coordinates of cells, respectively. (b)-(e) The pattern of particles at $t=5000$ for (b) $f=0.2$, $f_p=0.005$, (c) $f=0.2$, $f_p=0.1$, (d) $f=0.8$, $f_p=0.005$, and (e) $f=0.8$, $f_p=0.1$.}
\label{fig:rnd_dist}
\end{figure}

Let us examine how particles migrate depending oblateness $f$ and rotation parameter $f_p$. Figure~\ref{fig:rnd_dist} (a) shows an initial state of random distribution where particles are color-coded according to where they exist in the four regions. After a long enough time, the particles with small $f$ rarely move and do not mingle together (Figs.~\ref{fig:rnd_dist} (b) and (c)), while those with large $f$ actively construct between branches and are mixed (Figs.~\ref{fig:rnd_dist} (d) and (e)). Since mutual rotation is most active around $f_p=0.1$, it is seen that particles in the case Fig.~\ref{fig:rnd_dist} (e) do not produce thin branches.

\begin{figure}[h]
 \centering
 \begin{overpic}[width=16.0cm]{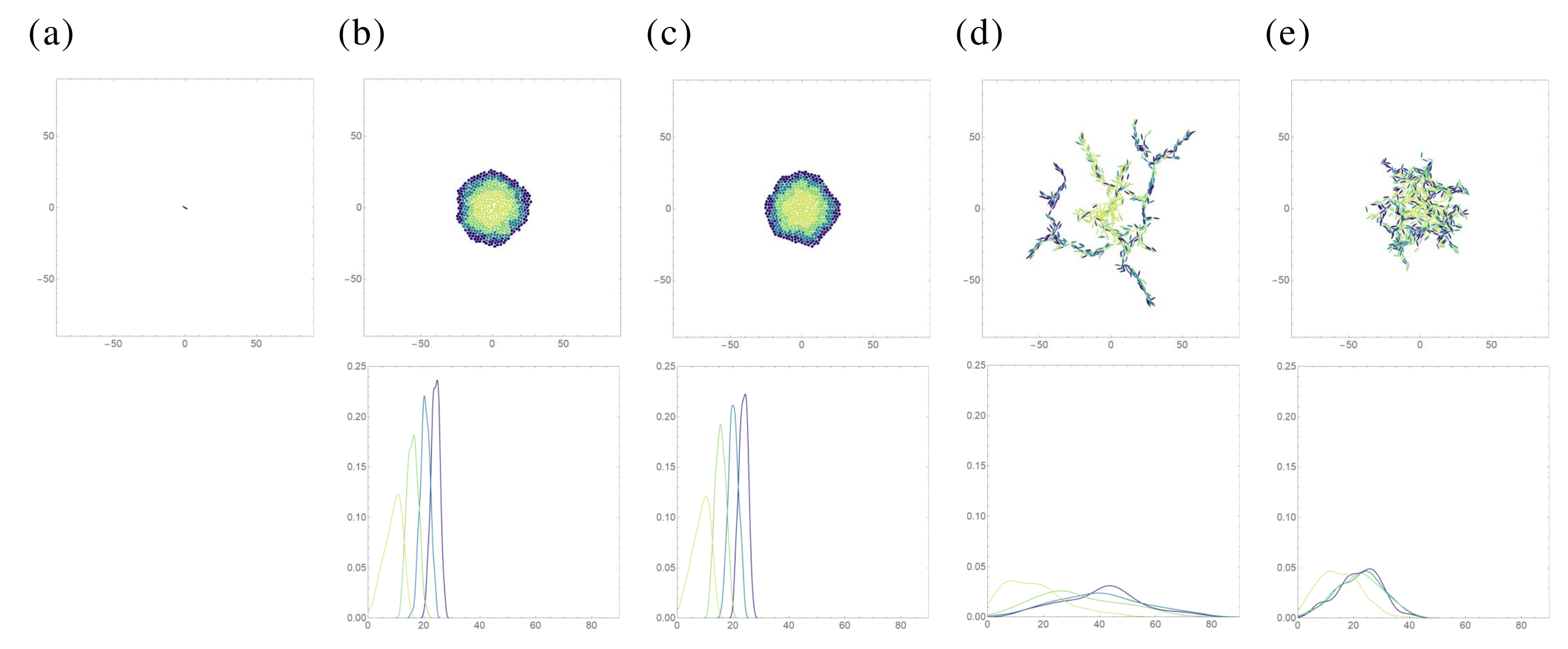}
 \end{overpic}
 \caption{Effect of oblateness $f$ and the rotation parameter $f_p$ on pattern in the case of constant supply. (a) At $t=0$, single particle is at the origin. The top figures of (b)-(e) show the patterns at $t=5000$ for (b) $f=0.2$, $f_p=0.005$, (c) $f=0.2$, $f_p=0.1$, (d) $f=0.8$, $f_p=0.005$, and (e) $f=0.8$, $f_p=0.1$.The bottom figures of (b)-(e) show the distribution of the four types of particles with respect to the distance from the origin. The horizontal axis represents the distance from the origin and the vertical axis represents density. The colors represent the order of the time step when a particle is supplied.}
\label{fig:cnt_dist}
\end{figure}

Figures~\ref{fig:cnt_dist} show the cell-mixing in the cases of constant supply. A particle is color-coded into four types according to the order of the time step when it is supplied. The upper figures of Figs.~\ref{fig:cnt_dist} (b)-(e) are the final patterns and lower figures show the distribution of the four types of particles with respect to the distance from the origin. The features of the patterns are similar to those in Figs.~\ref{fig:rnd_dist}.

\begin{figure}[h]
 \centering
 \begin{overpic}[width=14.0cm]{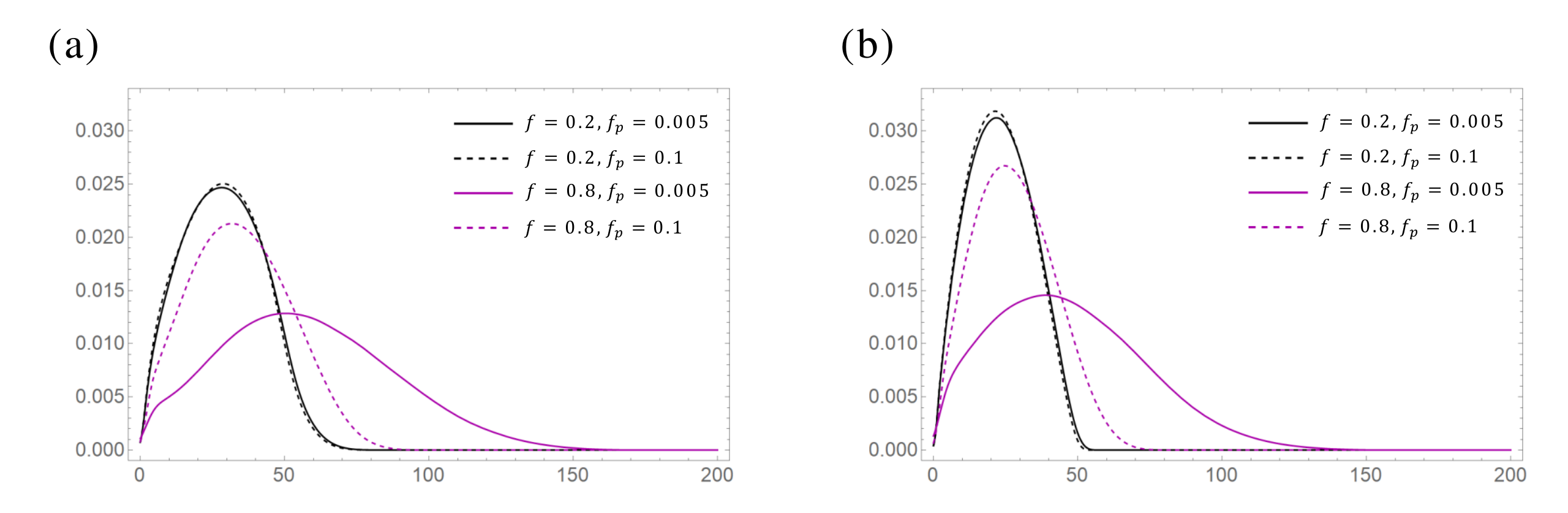}
 \end{overpic}
 \caption{Distribution of cell-to-cell distances for all pairs at $t=5000$ in the case of  (a) random distribution and (b) constant supply. The horizontal axis represents cell-to-cell distance and the vertical axis represents density.}
\label{fig:distances}
\end{figure}

Figure~\ref{fig:distances} (a) shows the distribution of distance between particles for Figs.~\ref{fig:rnd_dist} (b)-(e), and Fig.~\ref{fig:rnd_dist} (b) that for Figs.~\ref{fig:cnt_dist} (b)-(e). In both cases, the particles with large oblateness and small rotation parameter exhibit more widespread distributions.

\subsection{Boundary conditions are not essential for network structures}\label{sec:boundary}

\begin{figure}[h]
 \centering
 \begin{overpic}[width=13.0cm]{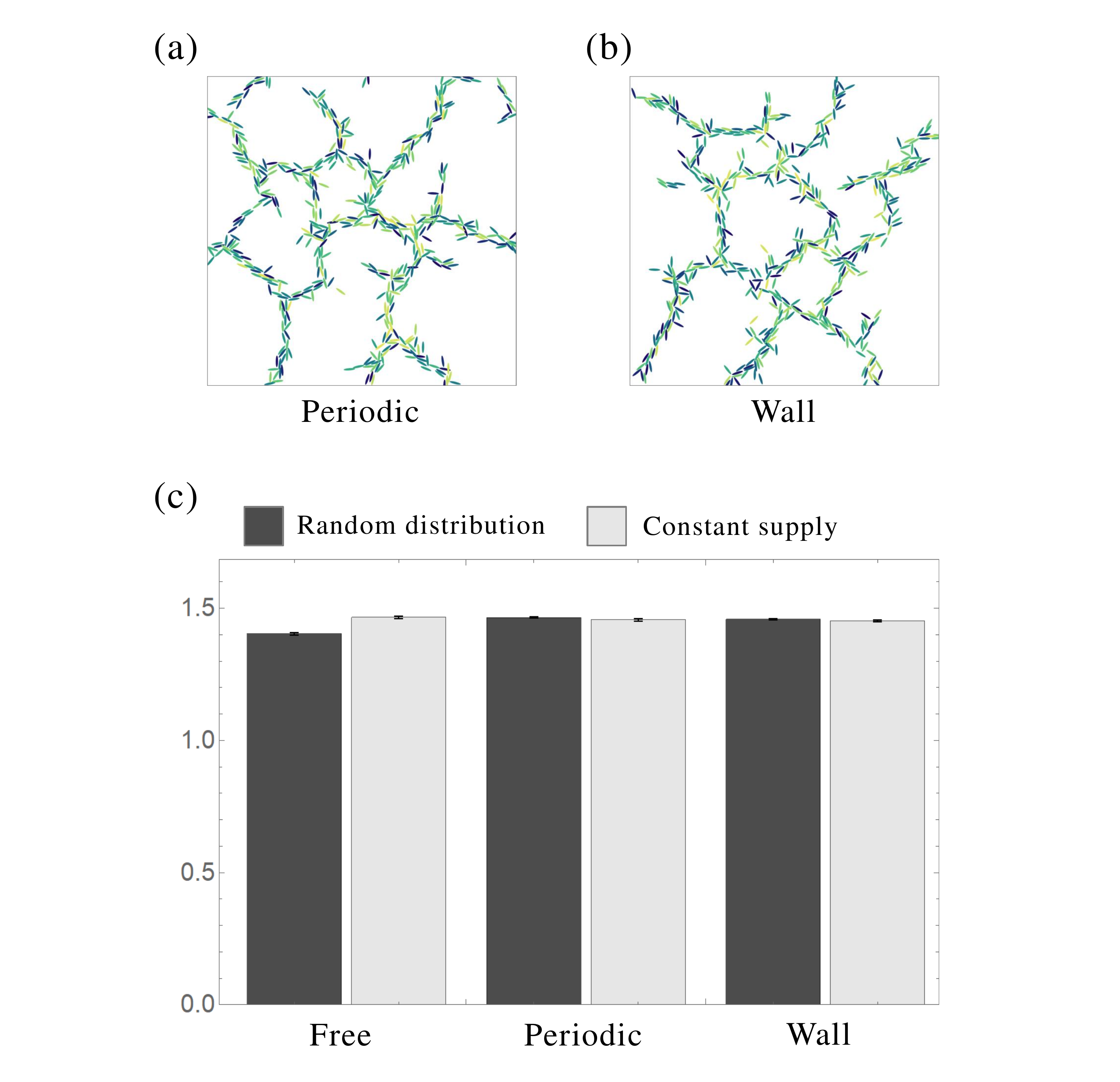}
 \end{overpic}
 \caption{The dependency of boundary conditions. The panels (a) and (b) show examples of simulation: (a) periodic boundary condition and (b) wall boundary condition (finite squared region). (c) Fractal dimension (Box-counting dimension) for three types of boundary conditions. Black bars show the result when ellipse particles are randomly placed as the initial configuration. Gray ones show the result when we supply an ellipse particle every ten steps to the origin. Parameters are $f=0.8$, $\gamma = 0.1$, $f_a = 0.002$, $f_r = 0.05$, and $f_p = 0.001$. The box is $[-120, 120]^2$ in periodic and wall boundary conditions. The data of (c) are averaged over $100$ simulations and their $95$\% confidence intervals.}
 \label{fig:boundary}
\end{figure}

We investigate the influence of boundary conditions on the pattern formation in our model. We consider a periodic boundary condition and a wall boundary condition (a finite square region). 
Particles with small oblateness or strong rotation form compact aggregates and hardly move from their initial positions. Their patterns are not affected by boundary conditions because branches do not elongate sufficiently. Thus, we examine only elongated particles with moderate value of the rotation parameter. Particles ($f=0.8$ and $f_p=0.001$) form branch-like structures in both periodic (Fig~\ref{fig:boundary} (a)) and wall (Fig~\ref{fig:boundary} (b)) boundary conditions. The fractal dimension remains almost unchanged irrespective of boundary conditions and simulation settings, as shown in Fig.~\ref{fig:boundary} (c).

%
%
%
%
%
%
\section{Concluding Remarks}
In this study, we have proposed a two-dimensional discrete mathematical model for angiogenesis, and investigated the pattern formation and the elongation of branches by numerical simulations. 
In the present model an EC is represented by an elliptic particle, and 
the dynamics of ECs is given by 
deterministic two-body interactions consisting of attraction and repulsion, and a rotation at contact. The attractive force represents a driving force 
induced by the contact between ECs through pseudopodia, and the repulsive force is due to the excluded volume effect. The rotation at contact is considered to be due to both excluded volume effect and reaction of pseudopodia. We have showed that the oblateness of an elliptic particle strongly affects the pattern formation. 
A similar model for angiogenesis with elliptic particles was considered in Ref.~\cite{palachanis2015}, where importance of the shape of particles was also demonstrated. Our numerical simulations have shown qualitatively 
oblateness dependence of patterns to theirs. Although our model is constructed from a different approach than theirs, it is suggested that the shape of a particle is essential to construct network structures. In addition, we have shown that the rotation of a particle also significantly affects pattern formation and the elongation of sprout. These results suggest that alignment of ECs by moderate rotation is essential to elongation of  branch structures.

Interestingly, even if the number of elliptic particles is relatively small, they align themselves in a straight line (Fig.~\ref{fig:10cells}). In the early stages of angiogenesis, a small number of vascular endothelial cells form sprouts in the extracellular matrix~\cite{arima2011}. By incorporating more realistic biological conditions such as the extracellular matrix 
, vascular endothelial growth factor, 
and three-dimensional elongation, we wish to improve the present model and quantitatively clarify the endothelial cell dynamics during angiogenesis so that we can use it for \textit{in silico} experiments in medical sciences.

%
%
%
%
%
%
%
\section*{Acknowledgements}
The authors would like to thank Dr. Kazuo Tonami and Mr. Kazuma Sakai for fruitful discussions about angiogenesis and mathematical modeling. TT is grateful for financial support to Arithmer Inc..
%
%
%
%
%
\bibliography{ellipse_model}
\bibliographystyle{plain}


%
\end{document}